\documentclass[conference]{IEEEtran}
\IEEEoverridecommandlockouts
\usepackage{cite}
\usepackage{amsmath,amssymb,amsfonts}
\usepackage{algpseudocode}
\usepackage{amsmath}
\usepackage{amssymb}
\usepackage{amsfonts}
\usepackage{xcolor}
\usepackage{hyperref}
\usepackage{tikz}
\usepackage{algorithm}
\usepackage{enumitem}
\usepackage{caption}
\usepackage{subcaption}

\usepackage{graphicx}
\usepackage{textcomp}
\usepackage{xcolor}
\usepackage{float}
\usepackage{booktabs}
\usepackage{array}
\usepackage{amsthm}
\theoremstyle{definition}
\newtheorem{definition}{Definition}[section]
\newtheorem{requirement}{REQ}[section]

\newcommand{\suchthat}{\, \mid \,}
\newcommand{\progressbar}[2]{%
    \begin{tikzpicture}[baseline=-0.1cm]
        \vspace{10mm}
        \fill[lightgray!30] (0,0) rectangle (1.5,0.3);
        \fill[blue!70] (0,0) rectangle ({1.5*#1/#2},0.3);
        \node[anchor=west] at (1.8,0.15) {\small\textcolor{black}{#1/#2}};
    \end{tikzpicture}%
}
\usepackage{listings, xcolor}

\definecolor{verylightgray}{rgb}{.97,.97,.97}

\lstdefinelanguage{Solidity}{
	keywords=[1]{anonymous, assembly, assert, balance, break, call, callcode, case, catch, class, constant, continue, constructor, contract, debugger, default, delegatecall, delete, do, else, emit, event, experimental, export, external, false, finally, for, function, gas, if, implements, import, in, indexed, instanceof, interface, internal, is, length, library, log0, log1, log2, log3, log4, memory, modifier, new, payable, pragma, private, protected, public, pure, push, require, return, returns, revert, selfdestruct, send, solidity, storage, struct, suicide, super, switch, then, this, throw, transfer, true, try, typeof, using, value, view, while, with, addmod, ecrecover, keccak256, mulmod, ripemd160, sha256, sha3}, 
	keywordstyle=[1]\color{blue}\bfseries,
	keywords=[2]{address, bool, byte, bytes, bytes1, bytes2, bytes3, bytes4, bytes5, bytes6, bytes7, bytes8, bytes9, bytes10, bytes11, bytes12, bytes13, bytes14, bytes15, bytes16, bytes17, bytes18, bytes19, bytes20, bytes21, bytes22, bytes23, bytes24, bytes25, bytes26, bytes27, bytes28, bytes29, bytes30, bytes31, bytes32, enum, int, int8, int16, int24, int32, int40, int48, int56, int64, int72, int80, int88, int96, int104, int112, int120, int128, int136, int144, int152, int160, int168, int176, int184, int192, int200, int208, int216, int224, int232, int240, int248, int256, mapping, string, uint, uint8, uint16, uint24, uint32, uint40, uint48, uint56, uint64, uint72, uint80, uint88, uint96, uint104, uint112, uint120, uint128, uint136, uint144, uint152, uint160, uint168, uint176, uint184, uint192, uint200, uint208, uint216, uint224, uint232, uint240, uint248, uint256, var, void, ether, finney, szabo, wei, days, hours, minutes, seconds, weeks, years},	
	keywordstyle=[2]\color{teal}\bfseries,
	keywords=[3]{block, blockhash, coinbase, difficulty, gaslimit, number, timestamp, msg, data, gas, sender, sig, value, now, tx, gasprice, origin},	
	keywordstyle=[3]\color{violet}\bfseries,
	identifierstyle=\color{black},
	sensitive=false,
	comment=[l]{//},
	morecomment=[s]{/*}{*/},
	commentstyle=\color{gray}\ttfamily,
	stringstyle=\color{red}\ttfamily,
	morestring=[b]',
	morestring=[b]"
}

\def\BibTeX{{\rm B\kern-.05em{\sc i\kern-.025em b}\kern-.08em
    T\kern-.1667em\lower.7ex\hbox{E}\kern-.125emX}}
\begin{document}

\title{Decoupling Reentrancy Protection from Smart Contract Implementation Logic
}

\author{
    \IEEEauthorblockN{Shashank Joshi}
    \IEEEauthorblockA{\textit{David R. Cheriton School of Computing} \\
    \textit{University of Waterloo}\\
    Waterloo, Canada \\
    s6joshi@uwaterloo.ca}
    \and
    \IEEEauthorblockN{Wojciech Golab}
    \IEEEauthorblockA{\textit{Electrical and Computer Engineering} \\
    \textit{University of Waterloo}\\
    Waterloo, Canada\\
    wgolab@uwaterloo.ca}
}

\maketitle

\begin{abstract}
Recently, DApps (Decentralized Applications) have witnessed exponential growth in their popularity across various sectors. Despite their seemingly robust inherent design, they suffer from a variety of vulnerabilities, which in turn can be exploited by an attack vector for gains. Among such vulnerabilities, reentrancy attacks are a persistent concern, with nefarious actors siphoning around 80M USD from the DApp ecosystem last year, leveraging the complex interaction associated with DApp and EVM’s (Ethereum Virtual Machine) message-passing semantics during inter-contract interaction. Existing research primarily focuses more on the detection of the reentrancy vulnerability and heavily relies on the existing patterns of attacks to identify the potential vulnerability and still fails to provide a precise solution to eliminate that loophole. Similarly, traditional reentrancy guards are ineffective for all the variations of reentrancy attacks, and the security guarantees are susceptible to the complexity of the interactions of the DApps. In this paper, we introduce a novel proxy-based approach designed to mitigate reentrancy vulnerabilities in smart contracts in a type-agnostic way. Unlike traditional reentrancy guards, our proposed solution integrates reentrancy logic directly into the proxy layer, intercepting all calls to the underlying implementation contract or set of contracts. Key features include a dual-mode operational system, offering both a gas-optimized internal guard and a high-security external lock registry for cross-contract reentrancy prevention. Furthermore, the proxy intelligently handles static calls, allowing for safe, non-blocking view-function execution while also protecting the implementation contract from ROR (Read Only Reentrancy) attacks. The intricate technicalities of the proposed proxy-based system, its implementation, and its application against reentrancy vulnerabilities are discussed within the scope of this paper. Through a rigorous analysis of its security coverage over existing solutions using a dataset of 70 vulnerable smart contracts, Sentinel achieves a 40\% higher success rate than existing methods, demonstrating superior coverage across four major reentrancy attack categories compared to existing solutions.
\end{abstract}


\section{Introduction}
Blockchain technology \cite{1, 2} has rapidly evolved, attracting widespread attention from government agencies, financial institutions, and tech companies \cite{3}. What began as the foundational layer for Bitcoin \cite{5} has since advanced into the Ethereum-driven 2.0 era \cite{4}, enabling the creation and deployment of smart contracts—self-executing code that automates agreement enforcement based on predefined conditions \cite{6, 7}. These smart contracts now underpin a wide range of decentralized applications (DApps), playing critical roles in financial systems, supply chains, and beyond, as evidenced by platforms like GSENetwork and Jingdong’s blockchain implementations \cite{9,10}.
With the rapid proliferation of decentralized applications (DApps) and the burgeoning value locked in smart contracts have amplified the criticality of smart contract security. Among the most insidious and historically damaging vulnerabilities is the reentrancy attack. Reentrancy is one of the vulnerabilities of Ethereum Virtual Machine (EVM) \cite{11} inter-contract semantics, where an external call to an untrusted contract is made before the calling contract's state variables are updated, allowing the untrusted contract to leverage this state inconsistency for repeated, unauthorized interactions. Several existing research pieces have demonstrated that a significant fraction of smart contracts deployed on the Ethereum blockchain exhibit programming errors, rendering them susceptible to potential reentrancy attacks \cite{12, 13}. This becomes starkly evident from a series of infamous attacks, such as the C.R.E.A.M. Finance attack, the DAO attack, the Fei-Rari attack, the Parity Wallet hack, etc., each resulting in millions of dollars in financial losses \cite{14}.
To address the threat of reentrancy attacks, current mitigation efforts encompass a range of solutions, including the introduction of safe code patterns, formal verification, migration towards safer programming languages, and symbolic execution. However, the significant majority of these efforts are concentrated towards identifying the potential vulnerabilities upfront in the development phase (or pre-deployment phase). Most of these detection tools \cite{17, 18, 19, 20} depend on the intermediate representation of vulnerable contracts and attack patterns to capture a viable attack path. However, this methodology results in suboptimal detection in more complex attack scenarios due to its restricted scope and rigid attack pattern design. Crucially, while these detection frameworks are valuable, they often fall short of providing concrete, deployable solutions to completely eliminate the identified loopholes, leaving a critical gap in the mitigation lifecycle. And addressing the vulnerabilities post-deployment is particularly challenging due to the following reasons:
\begin{enumerate}
    \item Immutable nature of smart contracts
    \item Anonymity of the owner
    \item Lack of coverage in offline static analysis and thus potential susceptibility to missing unknown run-time attack patterns.
\end{enumerate}
Furthermore, while there is a lack of reentrancy guards capable of thwarting reentrancy attacks (rather than merely detecting them), traditional reentrancy guards \cite{21} are typically limited in their scope and static in their application. Simple lock-based implementations, while preventing basic reentrancy, can inadvertently hamper the legitimate logic of the contract itself by overly restricting execution flows. Their effectiveness is highly contingent on meticulous implementation, and they become vulnerable if the attack path bypasses these existing, localized protections. Moreover, handling reentrancy across multiple related contracts or intelligently differentiating between malicious reentrant calls and legitimate static calls (which, despite their read-only nature, can still be part of a reentrancy chain if not properly accounted for) remains a complex and often unaddressed problem for standard implementations.
In this paper, we introduce Sentinel, a novel proxy-centric approach designed to comprehensively detect and deter reentrancy attacks. Unlike its conventional counterparts that reside within the application logic, Sentinel integrates reentrancy protection directly into the proxy layer of a potentially vulnerable contract. This strategic placement ensures that all calls to the underlying implementation contract are intercepted and protected before execution, inherently providing resilience to contract upgrades. The main contributions of this article are summarized as follows:
We provide a mathematical definition of reentrancy, precisely characterizing the vulnerability in terms of system states, execution traces, and critical state violations. This formalization offers an unambiguous foundation for analyzing and verifying reentrancy properties, moving beyond heuristic descriptions to enable more precise security reasoning and reentrancy guard design requirements.
\begin{itemize}
    \item We propose a dual-mode operational system, featuring a gas-optimized internal guard for efficient single-contract protection and a high-security external lock registry for robust, cross-contract reentrancy prevention. This architecture effectively defends against single-function, cross-function, and cross-contract reentrancy attacks.
    \item Our solution extends its protective scope to include view functions, specifically in the context of staticcall. A specialized check within the proxy detects and prevents ROR attacks, a subtle yet exploitable form of reentrancy, while simultaneously allowing legitimate read-only operations to proceed without unnecessary blocking.
    \item Our solution is engineered to be upgrade-proof. Upgrades or modifications to the smart contract protocol are confined to the implementation part of the contract.
    \item We also conduct a rigorous empirical evaluation of the Sentinel against a diverse dataset of 70 known vulnerable smart contracts.
    \item Our analysis includes a gas cost assessment and demonstrates superior security coverage across four major reentrancy attack categories compared to existing state-of-the-art mitigation techniques. Meanwhile, we also analyze the reasons and critical flaws associated with other reentrancy guards that rendered them ineffective against sophisticated reentrancy attacks.
\end{itemize}
The remainder of this paper is organized as follows: Section 2 provides some context or preliminary information on this work. Next, Section 3 analyzes and formalizes the reentrancy vulnerabilities. Section 4 presents a detailed description of our proposed model, its components and security approach under different reentrancy scenarios. Next, section 5 provides experimental results to evaluate Sentinel with other reentrancy guards. Section 6 discusses related works. Finally, section 7 concludes the paper by summarizing the contributions of our proposed model and identifying areas for future research and development.

\section{Background}
\subsection{Ethereum Accounts and Account Abstraction}
Ethereum has proposed Account Abstraction (AA) \cite{15} to address the intricacies surrounding user experience in blockchain interactions. This model aims to make blockchain technology more accessible to new users akin to Web2 and provide developers with manageable, modular components that facilitate easier troubleshooting and development \cite{16} while maintaining the principle of Web3 self-custody. Adoption of AA redefines the account paradigm by transitioning from the rigid structure of Externally Owned Accounts (EOAs), which are controlled by a single private key to enabling the smart contracts to function as primary user accounts called Contract Accounts (CAs). This allows for programmable, customizable logic that enhances security, usability, and flexibility. EOAs are the primary initiators of activity on the blockchain, capable of transferring ETH or other Ethereum-based tokens and initiating contract creation or function calls. They possess a nonce (transaction count) and a balance, but critically, they do not contain associated executable code. On the contrary, CAs are defined by their associated immutable bytecode and internal persistent storage. They are created by transactions originating from an EOA or another CA, and unlike EOAs, they are not controlled by a private key but by their embedded logic. CAs can hold Ethereum tokens or other Ethereum-based tokens and execute complex programs based on the required application logic. This ability to perform inter-contract message calls using various EVM opcodes makes CAs most relevant to reentrancy attack attack space.

\subsection{Ethereum Virtual Machine (EVM) Architecture}
The EVM is the runtime environment for smart contracts in the Ethereum ecosystem. It is a stack-based and quasi-turing complete (quasi because of its gas-bounded execution model that limits the number of execution processes to a finite number of computational steps) that executes bytecode instructions. Bytecodes is the compiled form of smart contracts, where each byte (or sequence of bytes) represents an opcode. These opcodes are the actual low-level executable instructions that the EVM processes using its stack-based architecture. Every node on the Ethereum network runs an instance of the EVM, ensuring that contract execution is deterministic and universally verifiable. The EVM through opcodes manages the global state of the blockchain, including account balances, contract code, and contract storage. During the execution of smart contracts, it maintains a call stack—when external calls occur, new frames are added, creating nested execution contexts for complex contract interactions. This peculiar mechanism along with the inter-contract message semantics creates a window of opportunity for the threat vectors to attempt a reentrancy attack to exploit this inconsistency in the states.

\subsection{Smart Contracts}
Smart contracts are the central component of the DApps and are responsible for storing and executing the business logic of that DApp. Introduced prior to the invention of blockchain and almost concurrently with the development of the contemporary Internet in \cite{22}, the concept remained dormant until the advent of Ethereum. They enabled the blockchain ecosystem to transition from a simple double-spending-resilient payment system to an efficient distributed ledger technology that can successfully emulate real-world interactions in a trustless and decentralized setting. Unlike traditional contracts, when particular conditions are met, these programs are developed to independently execute a particular set of instructions. These self-executing contracts enable parties to conduct business without the involvement of intermediaries, like in centralized paradigms. As described earlier, smart contracts in the Ethereum ecosystem are written in a high-level language such as Solidity. To execute in EVM, the smart contract is compiled into a bytecode, which, once deployed, is immutable and visible to all the users across the blockchain network.

\subsection{Types of Calls and Call Stack Management}
Interaction between accounts (CAs and EOAs) in Ethereum is performed through various types of call opcodes, each with a distinct set of properties regarding execution context, gas and state modification. EVM has 3 call opcodes which are summarized as:
\begin{itemize}
    \item call: call is the most common low-level function used to invoke functions in other contracts or send Ether. When this call opcode is invoked, the called contract operates in its own storage context and modifies the state of the same. As this opcode forwards all the gas to the called address, this provides an opportunity to perform a reentrancy attack. Reentrancy attackers often exploit call operations that occur before critical state update.
    \item delegatecall: This low-level call is used in proxy patterns. When used it executes code from another contract but using the calling contract’s context for instance If contract A executes a delegatecall to contract B, B’s code is executed in the storage context of A, which implies B can directly read from and write to A’s storage, and msg.sender and msg.value retain their original values from the initial call to A.
    \item staticcall: Introduce in EIP-214, this opcode is similar to call but enforces an extra constraint where any attempt to modify the blockchain state (opcodes such as SSTORE, CREATE, LOG) during staticcall or any subsequent nested call triggered by it will result in an immediate transaction REVERT (no gas forwarding). This guarantees read-only execution, making it suitable for querying immutable data or view functions. However, staticcall is susceptible to ROR attacks. Unlike other reentrancy attacks, ROR attacks indirectly exploit stale state information by a concurrent state-changing operation, potentially influencing external logic or decision-making processes.
\end{itemize}

\section{Reentrancy Attack Model}
In this section, we first discuss various types of reentrancy attacks, followed by a comprehensive framework for formally representing and analyzing these reentrancy vulnerabilities. Finally, we propose essential requirements for the reentrancy guard to effectively protect the victim smart contract from these vulnerabilities.
\begin{algorithm}
\caption{Victim: Reentrant Smart Contract}
\begin{lstlisting}[language=Solidity]
pragma solidity ^0.8.0;
contract Victim{
   mapping (address=>uint256) balances;
   function deposit() public payable{
       require(msg.value>0, "Please deposit some ETH");
       balances[msg.sender] += msg.value;
   }
   function withdraw() public{
       uint256 bal = balances[msg.sender];
       require(bal>0, "the user did not deposit that amount in this contract");
       (bool sent, ) = msg.sender.call{value: bal}("");
       require(sent, "Failed to send Ether");
       balances[msg.sender] = 0;
   }
}
contract Attacker{
   Victim public victim;
   constructor(address _victim) {
       victim = Victim(_victim);
   }
   receive() external payable{
       if(address(victim).balance>1 ether){
           victim.withdraw();
       }
   }
   function attack() external payable {
       require(msg.value > 0, "Send the required attack amount");
       victim.deposit{value: msg.value}();
       victim.withdraw();
   }
   function withdraw() public{
       (bool sent, ) = msg.sender.call{value: address(this).balance}("");
       require(sent, "Failed to withdraw Ether");
   }
}
\end{lstlisting}
\end{algorithm}
\subsection{Taxonomy of Reentrancy Attacks} \label{TRA}
At its core, reentrancy is a malicious exploit which leverages the EVM call stack mechanism where a smart contract's state-modifying logic is not completed before it makes an external call to an untrusted address which then makes a recursive call back to the original contract. Because the victim contract’s state has not yet been updated to reflect the initial operation, the re-entrant call can exploit the inconsistent state which can lead to severe consequences including but not limited to double spending, repeated withdrawals, or other forms of state corruption that are dependent on the previous stale state. The infamous DAO hack serves as the quintessential example of a reentrancy attack, where an attacker repeatedly drained funds before the contract could update its internal balances. The generalized code-level example for this vulnerability is demonstrated in Algorithm 1. There are different variations of reentrancy attacks depending upon their nature and target and it can be summarized as \cite{23}:

\subsubsection{Single-Function Reentrancy} 
\begin{figure}[h] 
   
    \centering
    \begin{center}
    \includegraphics[width=0.46\textwidth]{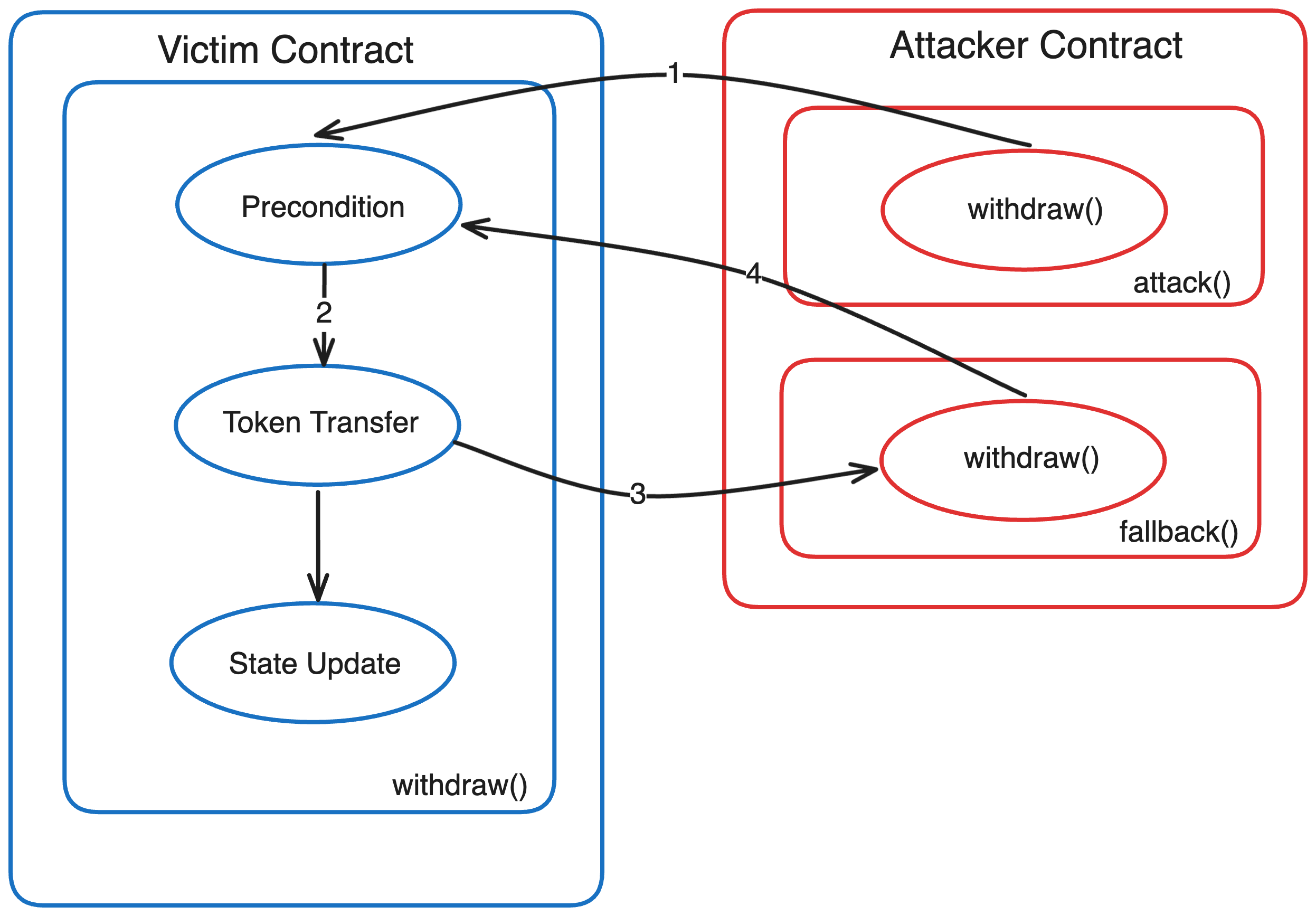}
    \end{center}
    \caption{An Example for Single Function Reentrancy Attack}
     \label{sfr}
\end{figure}

This is the most basic form of reentrancy attack. It occurs when a vulnerable function in a victim contract calls an external function, and that external function performs a recursive call that targets the same function that initiated the external call before the state of the first transaction is updated. Take the attack scenario illustrated in Figure \ref{sfr}, The attacker’s contract calls the withdraw() function of the victim contract. Subsequently, The victim contract performs a precondition check (such as require(balance[msg.sender] > 0)) and invokes an external call to transfer tokens to the attacker. This call automatically triggers the attacker's fallback function. The attacker's fallback function calls withdraw() on the victim contract again. Because the victim contract's state update is delayed until after the transfer, the attacker's balance is not yet zero, and the precondition check at step 2 passes, allowing the attacker to repeat the process by exploiting the inconsistent state.
\subsubsection{Cross-Function Reentrancy} 
\begin{figure}[h] 
   
    \centering
    \begin{center}
    \includegraphics[width=0.46\textwidth]{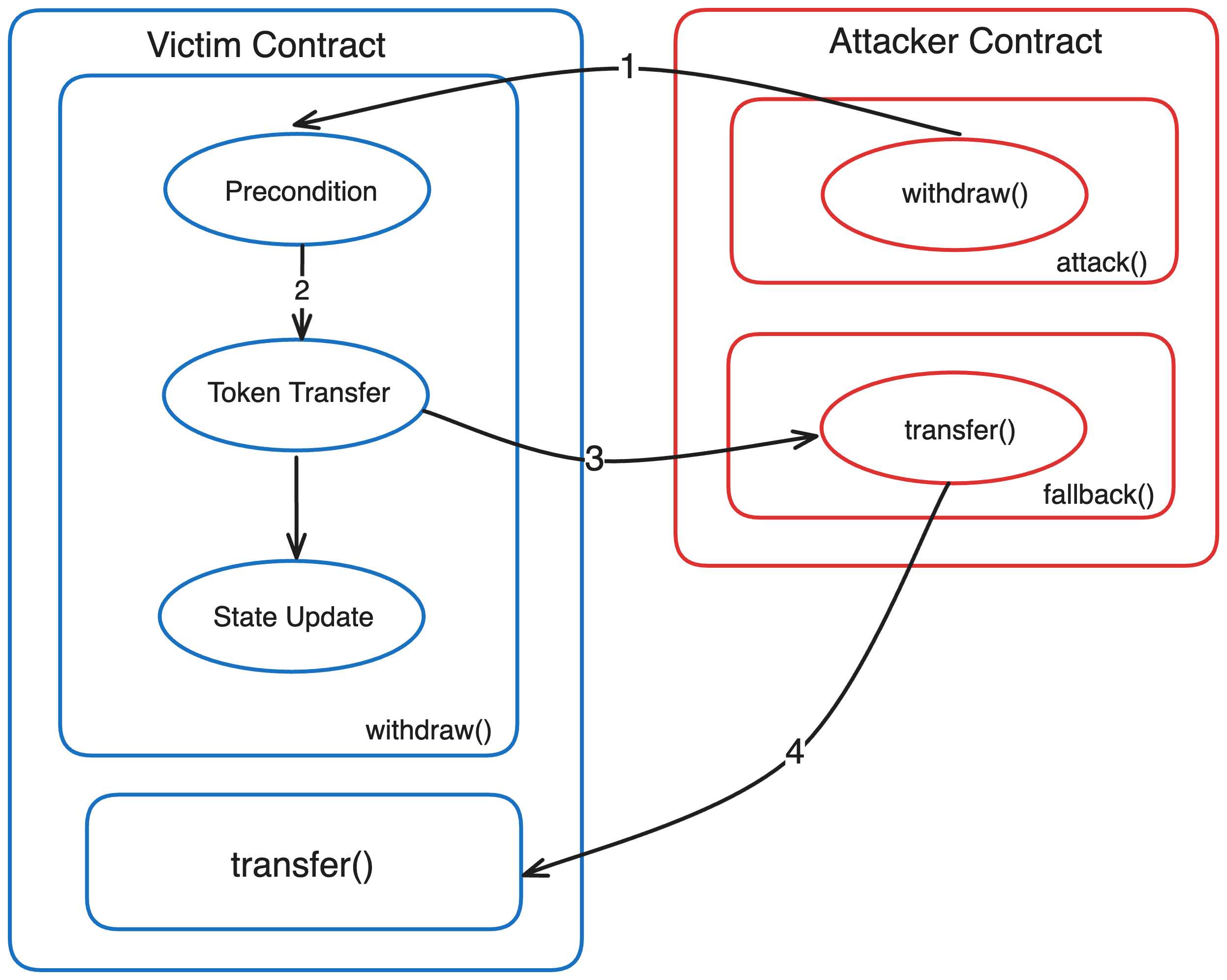}
    \end{center}
    \caption{An Example for Cross Function Reentrancy Attack}
     \label{cfr}
\end{figure}

In this variant, the attacker exploits two different functions of the same contract. The entry function call sets up the state change, and the re-entrant call made through a different function exploits the delayed state change initiated by the entry function. This type of attack is more sophisticated as it involves manipulating the interdependencies between different functions. Figure \ref{cfr} highlights a common cross-function reentrancy scenario where the attacker calls withdraw() function on the victim contract, to withdraw the associated balance from the same. The victim contract checks the preconditions and invokes an external call to transfer tokens to the attacker. Now this is the crucial phase where the cross-function branches away from the single function reentrancy, instead of entering the same entry function, the attacker's fallback() function re-enters the victim contract through transfer() that has a shared state with the entry function.
\subsubsection{Cross-Contract Reentrancy} 
\begin{figure}[h] 
     
    \centering
    \begin{center}
    \includegraphics[width=0.46\textwidth]{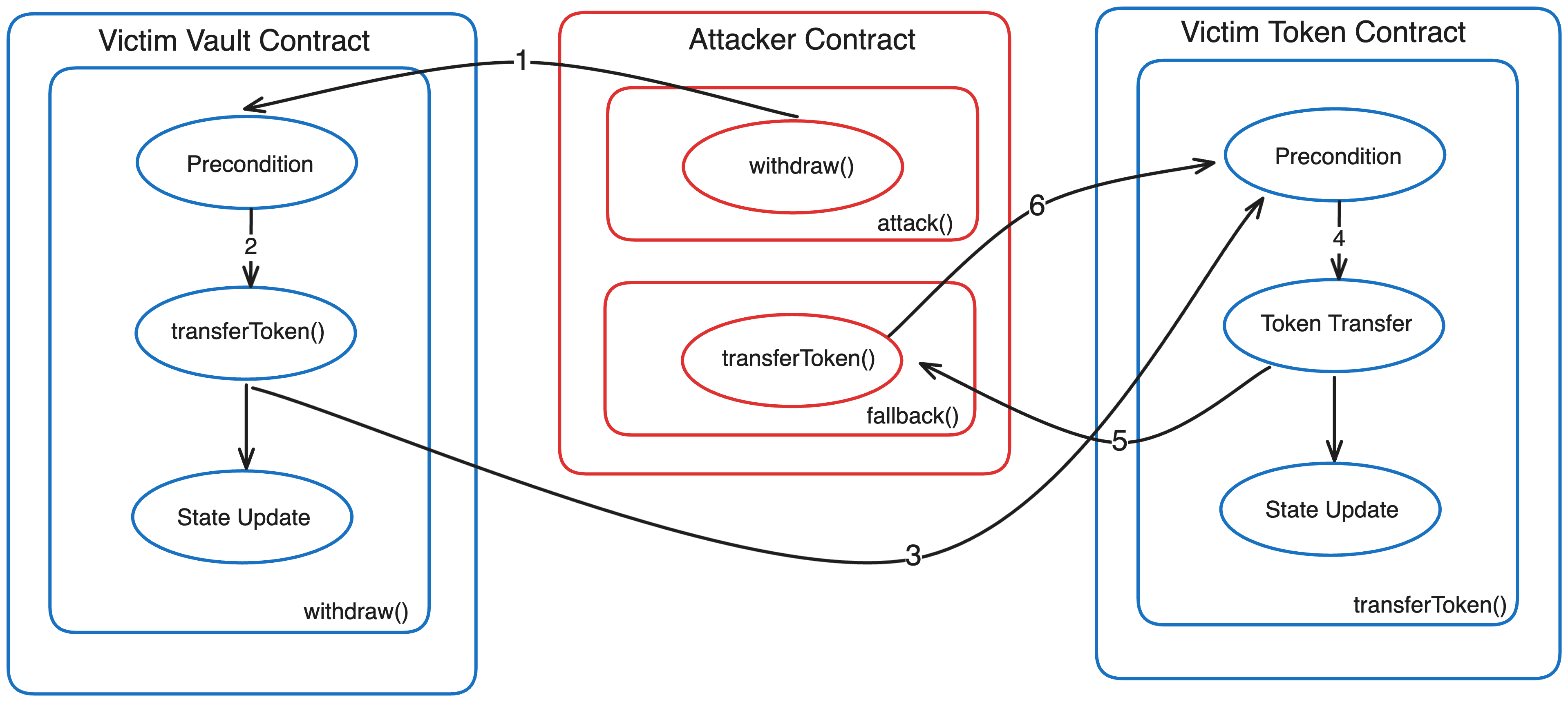}
    \end{center}
    \caption{An Example for Cross Contract Reentrancy Attack}
    \label{ccr}
\end{figure}

This complex attack involves a malicious interaction between multiple contracts, typically the attacker, the victim, and another legitimate contract (e.g., a token contract). As shown in Figure \ref{ccr}, the attack is initiated by the attacker calling a function on the Victim Vault Contract, which subsequently makes an external call to a separate Victim Token Contract to transfer funds. This transfer, in turn, triggers the attacker's fallback function, which re-enters the Victim Token Contract. The attacker's re-entrant call exploits an incomplete state or race condition across the multiple contracts, allowing for unintended state manipulation or unauthorized access.
\subsubsection{Read-Only Reentrancy} 
\begin{figure}[h] 
     
    \centering
    \begin{center}
    \includegraphics[width=0.46\textwidth]{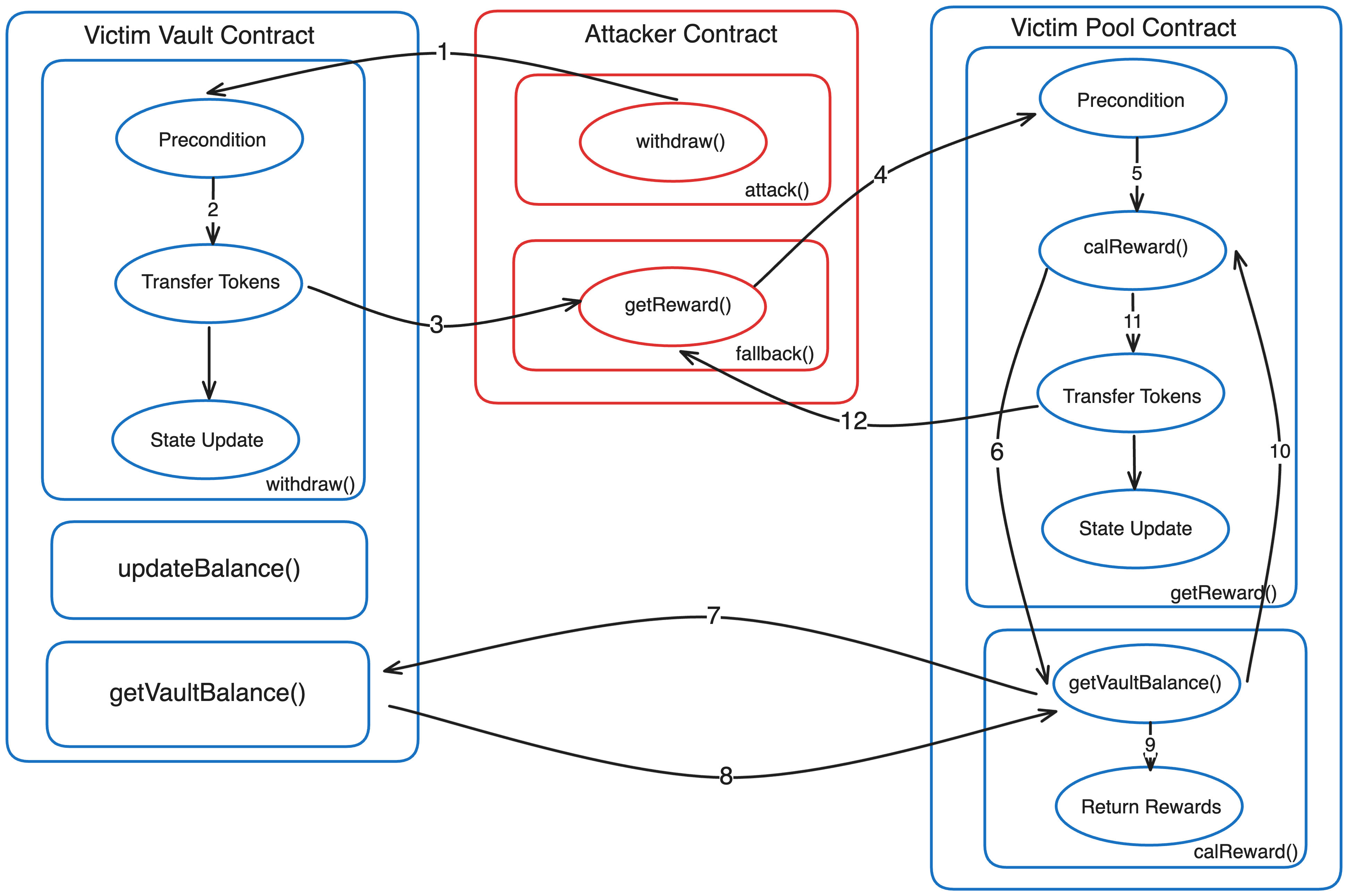}
    \end{center}
    \caption{An Example for Read Only Reentrancy Attack}
    \label{ror}
\end{figure}

A more subtle form of cross-contract or cross-function reentrancy, this attack involves view or pure functions, which are read-only and do not alter the state. An attacker could use these functions to gather information or set conditions favourable for a subsequent reentrancy attack. While these attacks don’t directly drain the funds from the victim contracts, they can be part of a larger strategy that leads to favourable outcomes for the attacker. A clear attack pattern is highlighted in Figure \ref{ror} where the attacker initiates a call to the victim vault contract's withdraw() function, which triggers the attacker's fallback() function, which then calls another contract (the victim pool contract) to calculate a reward. Here, the victim pool contract queries the victim vault contract's state (e.g., its balance) to perform its calculation. The external call to transfer the calculated rewards triggers the attacker's fallback function. This is where the re-entrant call is made, as the attacker's fallback function re-enters the Victim Vault Contract to read its state before it has been fully updated. The attacker can then exploit this outdated information, leading to inflated rewards.
\subsection{Formal Model for Reentrancy Attack}
Having established the taxonomy of reentrancy attacks and their characteristic patterns, we now propose a comprehensive mathematical framework that transcends pattern-based heuristics to formalize the fundamental problem space of reentrancy vulnerabilities and establish the key requirements for prevention mechanisms. To formalize reentrancy vulnerabilities, we first establish a foundational model for Solidity smart contracts. Let $C$ be a smart contract instance residing on-chain, characterized by its immutable code and mutable storage. We define its state space as:\\
\begin{itemize}
    \item $S = \{ \sigma_1, \sigma_2, ....., \sigma_N \}$: a finite set of all possible contract states, where $N$ is the total number of reachable states.
    \item $\sigma \in S$:the current contract state, comprising the values of all its storage variables, balances (ETH and associated tokens).
    \item $F =\{f_1, f_2,...,f_M\}$: a finite set of public and external functions defined within contract $C$.
    \item $T$: a sequence of discrete time steps, representing the sequential execution of transactions.
\end{itemize}
Similarly, a function $f \in F$ can be modeled as a state-transition system as:
$$f: S \times Args \rightarrow S' \times Return \times Effects$$
where:\\
$Args$: a tuple of input arguments supplied to the function.\\
$Return$: a tuple of values returned by the function.\\
$Effects$: a set of side effects generated by the function's execution such as external calls to external addresses.\\
Also, Let $E(C, f, \sigma, t)$ denote the execution of a function $f$ starting in state $\sigma$ at time $t$ with context to contract $C$ with final state as $\sigma_{final}$. Finally, $Completed(C,f,t_i,t_f)$ denotes whether the function $f$ of Contract $C$ that initiated at time $t_i$ is still active in EVM call stack at time $t_f$.

As we have observed in reentrancy attack patterns in section \ref{TRA}, reentrancy vulnerability exists in a smart contract $C$ if there exists a sequence of operations such that a function $f \in F$ or a related function within $C$ or a related contract $C’$ is re-executed, directly or indirectly, before its initial invocation has completed its intended state modifications, leading to an inconsistent or exploitable state. Building upon this, we can formalize the different reentrancy attacks as (Note: for all the reentrancy scenarios the execution of the entry function is triggered by the attacker):
\begin{definition}[Single Function Reentrancy (SFR)]
SFR occurs in contract $C$ through function $f \in F$ if, during an execution path of $f$ from state $\sigma_1$ at time $t_1$, an external call is made, which subsequently causes $f$ itself to be re-entered (recursively invoked) in an intermediate state $\sigma_2$ (where $\sigma_2 \neq \sigma_{final}$, indicating unfinalized state changes), leading to an exploit.
\begin{align}
SFR(C,f) &\implies \exists \sigma_1, \sigma_2 \in S \exists t1<t2<t3 \in T \nonumber \\
&\quad \suchthat E(C,f, \sigma_1, t_1) \nonumber \\
&\quad \rightarrow_{ExtCall(C_{attacker}, f')} \nonumber \\
&\quad \rightarrow E(C_{attacker},f', \sigma', t_2) \nonumber \\
&\quad \rightarrow_{ExtCall(C, f)} \nonumber \\
&\quad \rightarrow E(C,f, \sigma_2, t_3)
\end{align}
where:
$$\sigma_2 \neq \sigma_{final} \wedge \neg Complete(C, f, t_1, t_3)$$
\end{definition}
\begin{definition}[Cross Function Reentrancy (CFR)]
CFR occurs in contract $C$ through function $f_1, f_2 \in F$ where $f_1 \neq f2$,  if an external call initiated by $f_1$ allows $f_2$ to be called back into $C$, and $f_1$ and $f_2$ operate on shared critical state variables in a way that the re-entry of $f_2$ exploits $f_1$'s unfinalized state changes.

\begin{align}
CFR(C,f_1, f_2) &\implies \exists \sigma_1, \sigma_2 \in S \exists t1<t2<t3 \in T \nonumber \\
&\quad \suchthat E(C,f_1, \sigma_1, t_1) \nonumber \\
&\quad \rightarrow_{ExtCall(C_{attacker}, f')} \nonumber \\
&\quad \rightarrow E(C_{attacker},f', \sigma', t_2) \nonumber \\
&\quad \rightarrow_{ExtCall(C, f_2)} \nonumber \\
&\quad \rightarrow E(C,f_2, \sigma_2, t_3)
\end{align}
where:
$$SharedState(f_1,f_2)\wedge\sigma_2 \neq \sigma_{final} \wedge \neg Complete(C, f_1, t_1, t_3)$$
And,
$$SharedState(f_1,f_2) \implies \exists v \in StorageVars(C)$$
$$\suchthat (Modifies(f_1,v)\vee Reads(f_1,v))$$
$$\wedge (Modifies(f_2,v)\vee Reads(f_2,v))$$
\end{definition}

\begin{definition}[Cross Contract Reentrancy (CCR)]
CCR occurs in contracts $C$, $C’$ involving function $f_1 \in F$ and $f'_1 \in F’$ respectively, if $f_1$ makes state modifications followed by an external call to $C’$, followed by an external call made by $f'_1$, which subsequently leads to a re-entry in $C’$, exploiting an unfinalized state in $C’$.

\begin{align}
CCR(C, C',f_1, f'_1) &\implies \exists \sigma'_1, \sigma'_2 \in S' \exists t1<t2<t3<t4 \in T \nonumber \\
&\quad \suchthat E(C,f_1, \sigma_1, t_1) \nonumber \\
&\quad \rightarrow_{ExtCall(C', f'_1)} \nonumber \\
&\quad \rightarrow E(C',f'_1, \sigma'_1, t_2) \nonumber \\
&\quad \rightarrow_{ExtCall(C_{attacker}, f')} \nonumber \\
&\quad \rightarrow E(C_{attacker},f', \sigma', t_3) \nonumber \\
&\quad \rightarrow E(C',f'_1, \sigma'_2, t_4)
\end{align}
where:
$$\sigma'_2 \neq \sigma'_{final} \wedge \neg Complete(C', f'_1, t_2, t_4)$$
\end{definition}
\begin{definition}[Read Only Reentrancy (ROR)]
ROR occurs in contract $C$ involving function $f_1\in F$ and $f_2\in F$ (or a function from other contract $C’$), if $f_1$ makes state modifications followed by an external call, and during the external call, an attacker or another contract (invoked by the attacker) can observe the intermediate, unfinalized state of $C$, leading to a decision or action that is detrimental to $C$ when $f_1$ eventually completes.

\end{definition}

Based on the formal models for these reentrancy attacks, to be effective reentrancy guard the solution should satisfy the following requirements:

\begin{requirement}[Atomicity Preservation]
If an external call occurs at time $t_c$, then no state modification should occur at any later time $t_m>t_c$ during the execution $f$.
\end{requirement}
\begin{requirement}[Static Protection for Inconsistent State Exposure]
Any view into a contract’s state from other contracts (or EoAs )must only see valid, invariant-holding states, even if the current contract is in the middle of a transaction.
\end{requirement}
\begin{requirement}[Scope-Aware State Coupling ]
For $C = \{C_1, C_2, ....., C_n\}$ a finite coupled set of contracts (sharing state logic or invariants), the above two requirements should hold at the domain level and not at the contract level.
\end{requirement}

\section{Proposed solution} \label{psol}
This section presents a detailed technical description of our proposed Sentinel model, its architectural design, core components, and the flow of execution. Our approach integrates a dual-mode reentrancy guard directly into the proxy layer ensuring that reentrancy protection is decoupled from the application logic, providing a gas-efficient, upgrade-resilient, and highly configurable security layer against various reentrancy attack vectors. 
  \begin{figure}[h] 
    \centering
    \begin{center}
    \includegraphics[width=0.46\textwidth]{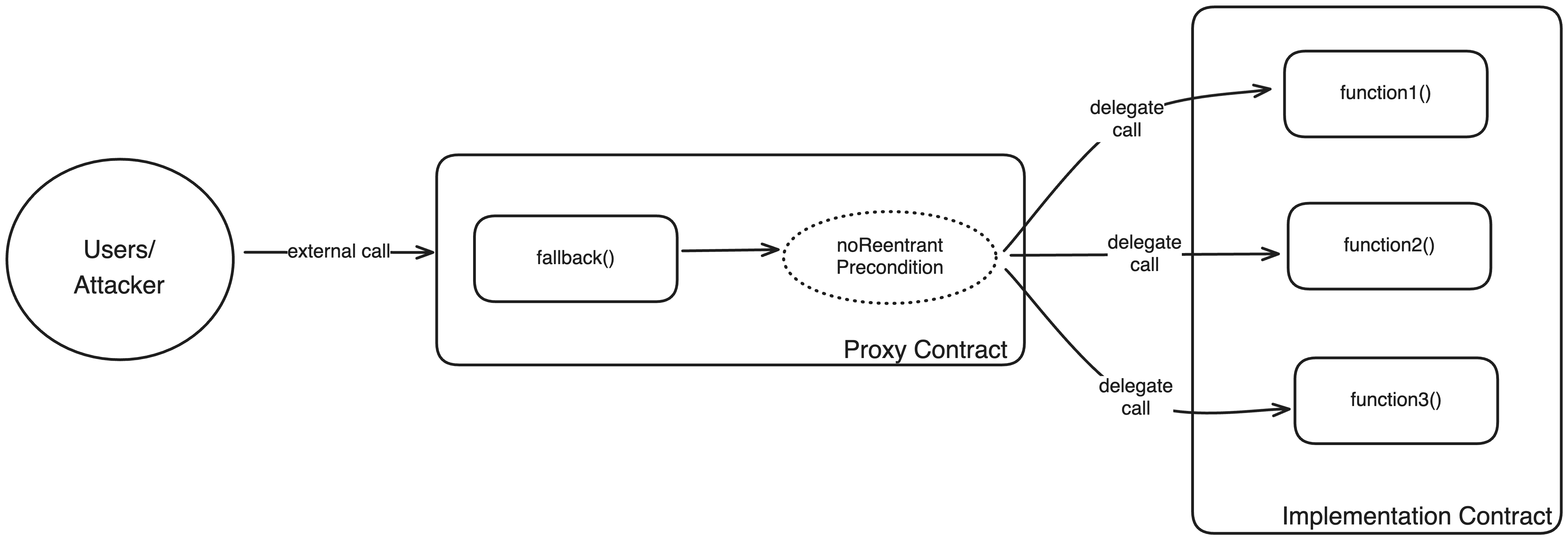}
    \end{center}
    \caption{Architecture Diagram For Sentinel Optimized Mode}
\end{figure}
\subsection{Architectural Overview}
The Sentinel establishes a flexible security architecture by acting as a single, immutable, transparent entry point for all the interactions with an underlying logic (implementation) contract. As depicted in Figure 5, Sentinel establishes a clear separation of concerns where the proxy is solely responsible for security and upgradability, while the implementation contract focuses on the core business logic.
The key components of our model are:
1. SentinelProxy: The user-facing entry point. It encapsulates the entire reentrancy guard logic, manages the implementation address, and handles administrative functions (e.g., upgradeTo(), setMode()). All user interactions flow through its fallback() or receive() functions.
2. ILockRegistry: An interface implementation of LockRegistry that provides a global, shared locking mechanism. In the Sentinel's high-security mode, it interacts with this registry via standard calls to lock(), unlock(), and isLocked() functions, enabling reentrancy protection that spans across multiple proxy instances or logical domains.
3. Implementation Contract(s): These are the contracts containing the DApp's core business logic. They are unaware of the proxy and are designed as standard Solidity contracts. The proxy's delegatecall ensures their code executes against the proxy's context.
\subsection{Core Component and Mechanism}
\subsubsection{The SentinelProxy Contract} \label{pcon}
Our proxy leverages the fallback() function [a special solidity procedure that acts as a catch-all for incoming transactions invoked without calling any specific function] to intercept all incoming calls, regardless of their signature or whether they contain data. This universal interception is crucial, as it ensures that every execution path to the implementation contract is first subjected to our reentrancy check. Upon interception, the proxy's internal control flow executes first before forwarding the original call's calldata to the designated implementation contract using the delegatecall. The proxy’s primary logic, as shown in Algorithm \ref{alg:main_execution_flow}, is structured as:
Pre-Execution Phase: An initial security check that verifies whether the current call is a legitimate entry point or a reentrant attempt. This phase identifies whether the incoming call is a staticcall or an external call, and performs the required checks based on the security mode selected for the concerned setup.
Call Forwarding: If the check passes in the pre-execution phase, the call data and context are forwarded to the implementation contract using the delegatecall. This ensures that the implementation’s code is executed in the proxy's storage context, preserving the state of the proxy and enabling upgradeability.
Post-Execution Cleanup: After the delgatecall returns, the proxy's state is reset to allow for future calls.
\begin{algorithm}[H]
    \caption{\texttt{SentinelProxy} Main Execution Flow}
    \label{alg:main_execution_flow}
    \small
    \begin{algorithmic}[1]
        \Procedure{fallback}{} 
           
            \State \texttt{isStaticCall} $\gets$ \Call{\texttt{\_isStaticCall}}{}
            \State
            
            \If{NOT \texttt{isStaticCall}}
                
                \If{\Call{GetBool}{\texttt{MODE\_SLOT}} is TRUE (High-Security Mode)}
                    \State \Call{\texttt{\_highSecurityGuard}}{} 
                \Else 
                    \State \Call{\texttt{\_optimizedGuard}}{} 
                \EndIf
            \Else  is TRUE
                
                \State \textbf{Require} \Call{GetBool}{\texttt{STATIC\_CALL\_ALLOWED\_SLOT}} is TRUE, "Reentrancy Error (Static Mode)"
            \EndIf
            \State
            
            \State \texttt{implAddress} $\gets$ \Call{GetAddress}{\texttt{IMPLEMENTATION\_SLOT}}
            \State \texttt{calldata} $\gets$ Current Call Data
            \State \texttt{gasRemaining} $\gets$ Current Gas
            \State \texttt{result}, \texttt{returndata} $\gets$ \Call{ExecuteDelegateCall}{\texttt{implAddress}, \texttt{calldata}, \texttt{gasRemaining}}
            \State
            
            \If{NOT \texttt{isStaticCall}}
                \If{\Call{GetBool}{\texttt{MODE\_SLOT}} is TRUE (High-Security Mode)}
                    \State \Call{\texttt{\_unlockHighSecurity}}{} 
                \Else 
                    \State \Call{SetUint}{\texttt{LOCK\_STATUS\_SLOT}, 0} 
                    \State \Call{SetBool}{\texttt{STATIC\_CALL\_ALLOWED\_SLOT}, true} 
                \EndIf
            \EndIf
            \State
             result
            \If{\texttt{result} is successful}
                \State \textbf{Return} \texttt{returndata}
            \Else
                \State \textbf{Revert} with \texttt{returndata}
            \EndIf
        \EndProcedure
    \end{algorithmic}
\end{algorithm}
\subsubsection{Dedicated Storage Slot Management}
In Solidity, contract storage is organized as a contiguous array of 32-byte slots (starting from slot 0). Variables are allocated to these slots sequentially based on their declaration order and packing rules. In a proxy pattern, where an implementation contract's code executes in the storage context of the proxy, a critical challenge of storage collision arises. If the proxy and the implementation both declare variables at the same storage slot, their data would overwrite each other, leading to unpredictable behaviour and potential vulnerabilities. In Sentinel, because the consistency of the proxy’s state is very essential to enforce the reentrancy protection therefore to circumvent potential storage collision, Sentinel employs dedicated storage slots where instead of using sequential slots, which are likely to be occupied by implementation’s variable, we store the proxy’s critical configuration data in slots derived from the hash of unique and descriptive strings. The hash function generates a 32-byte digest from an arbitrary input. By using a hash of a unique string as a storage key, the resulting slot makes it improbable for an implementation contract to accidentally allocate a variable to the same high-entropy slot used by the proxy.
\subsubsection{LockRegistry}
LockRegistry and its interface ILockRegistry are pivotal to the functionality of Sentinel’s high-security mode( detailed in Section \ref{highsec}). As per identified Requirement 3, an idle reentrancy guard should be aware of the exact boundary of operations and Sentinel fulfills this requirement through LockRegistry. This external contract provides a decentralized, global mutex mechanism that can span across multiple related smart contracts or even different logical domains within a larger DApp ecosystem. This contract maintains the locked status of each registered domain and address that currently holds the lock for a given domain.
\subsubsection{Static Call Detection and Protection}
\begin{algorithm}[H]
    \caption{\texttt{\_isStaticCall()} Detection}
    \label{alg:is_static_call}
    \begin{algorithmic}[1]
        \Function{\texttt{\_isStaticCall}}{} 
             \State \textbf{try:}
            \State \quad \texttt{AttemptStateChangeOperation()} \Comment{e.g., \texttt{log0()} in assembly}
            \State \quad \textbf{return} FALSE \Comment{Not a static call, as state change was possible}
            \State \textbf{catch} exception: \Comment{e.g., \texttt{STATICCALL} preventing state change}
            \State \quad \textbf{return} TRUE \Comment{It is a static call, as state change was reverted}
        \EndFunction
    \end{algorithmic}
\end{algorithm}
A critical feature of our design is its ability to differentiate and handle staticcall safely. As shown in Algorithm \ref{alg:is_static_call}, our proxy design includes a private function \_isStaticCall() designed to detect if the incoming call is a static context. The function leverages the fact that the EVM strictly enforces its read-only nature, causing a revert on state change. This function leverages a try-catch block that attempts to execute a minimal state-changing assembly instruction, specifically log0(0, 0). If the call is a staticcall, the log0 instruction triggers an immediate REVERT. This revert is caught by our try-catch block, allowing Sentinel to catch the staticcall. This detection mechanism is designed to incur minimal gas cost on revert, although REVERT does not refund gas, limiting the gas provided to the try block ensures that the maximum gas lost for a staticcall detection is capped at this very small, predefined amount (1000 gas), preventing significant overhead even when a revert occurs. This design choice to segregate the static call is deliberate and gas-efficient as static calls do not trigger the state-changing locks, saving the gas cost associated with SSTORE operations or external LockRegistry calls.
\subsection{Dual-Mode Reentrancy Guard}
Sentinel offers a dual-mode reentrancy guard, which facilitates flexible and adaptive security. This mode is selected via the MODE\_SLOT based on the requirements and scope of the contract domain.
\subsubsection{Optimized Mode}
As depicted in Figure 5, this mode provides a gas-efficient, single-contract reentrancy check. The mechanism relies on the LOCK\_STATUS\_SLOT, a uint256 storage variable within the proxy's dedicated storage.

When the \_optimizedGuard() function is invoked, refer to Algorithm \ref{alg:optimized_guard}, it first performs an SLOAD operation on LOCK\_STATUS\_SLOT. It then requires that the retrieved value is 0 (representing an unlocked state). If the value is 1 (indicating an active lock), the transaction immediately reverts with a "Reentrant call (optimized mode)" error, preventing further execution.

Immediately after the check passes, an SSTORE operation sets LOCK\_STATUS\_SLOT to 1. This status state change occurs before the delegatecall to the implementation contract, ensuring that the lock is established before any external (re-entrant capable) interaction occurs. Concurrently, the STATIC\_CALL\_ALLOWED\_SLOT is set to false to mitigate ROR attacks during the locked state. This is followed by the execution of the implementation logic in the proxy's context and return of the delgatecall. Upon the successful return of the delgatecall from the implementation contract, the proxy's fallback() function executes a post-call cleanup as described in section \ref{pcon} where another SSTORE operation resets LOCK\_STATUS\_SLOT back to 0, re-opening the lock for subsequent legitimate calls. The STATIC\_CALL\_ALLOWED\_SLOT is also reset to true.

While highly effective and gas-efficient for preventing re-entry into the same proxy instance, this mode's lock is local to the proxy's storage, thus it has limited scope (or domain). Hence, it does not inherently prevent re-entry attempts originating from other contracts that are not proxied by this specific instance, or re-entry into a different, unproxied contract within a broader execution domain. It is primarily designed for single-point-of-entry reentrancy protection.
\begin{algorithm}[]
    \caption{\texttt{\_optimizedGuard()} (Optimized Mode Logic)}
    \label{alg:optimized_guard}
    \begin{algorithmic}[1]
        \Procedure{\texttt{\_optimizedGuard}}{} \Comment{Internal function}
            \State \Comment{Check if the internal lock is already active}
            \State \textbf{Require} \Call{GetUint}{\texttt{LOCK\_STATUS\_SLOT}} is 0, "Reentrancy Error (Optimized Mode)"
            \State
            \Comment{Activate the internal lock}
            \State \Call{SetUint}{\texttt{LOCK\_STATUS\_SLOT}, 1}
            \State \Comment{Disallow static calls while the lock is active (prevents ROR)}
            \State \Call{SetBool}{\texttt{STATIC\_CALL\_ALLOWED\_SLOT}, false}
        \EndProcedure
    \end{algorithmic}
\end{algorithm}
\subsubsection{High-Security Mode} \label{highsec}
\begin{algorithm}[]
    \caption{\texttt{\_highSecurityGuard()} (High-Security Mode Logic)}
    \label{alg:high_security_guard}
    \begin{algorithmic}[1]
        \Procedure{\texttt{\_highSecurityGuard}}{} \Comment{Internal function}
            \State \texttt{domainId} $\gets$ \Call{GetBytes32}{\texttt{DOMAIN\_ID\_SLOT}}
            \State \texttt{lockRegistry} $\gets$ \Call{GetAddress}{\texttt{LOCK\_REGISTRY\_SLOT}}
            \State
            \Comment{Check if the external domain lock is already active}
            \State \textbf{Require} \texttt{lockRegistry.isLocked(domainId)} is FALSE, "Reentrancy Error (High-Security Mode)"
            \State
            \Comment{Activate the external domain lock}
            \State \texttt{lockRegistry.lock(domainId)}
            \State \Comment{Disallow static calls while the lock is active (prevents ROR)}
            \State \Call{SetBool}{\texttt{STATIC\_CALL\_ALLOWED\_SLOT}, false}
            \State \textbf{Emit} \texttt{ReentrancyLockActivated(domainId)}
        \EndProcedure
    \end{algorithmic}
\end{algorithm}

The high-security mode fills the loophole of the optimized mode by providing cross-contract reentrancy protection through integration with a LockRegistry contract. As discussed earlier, this architecture establishes a global, shared mutex system capable of coordinating reentrancy locks across multiple SentinelProxy instances and other compatible contracts within the execution domain, as shown in Figure 6.
  \begin{figure}[H] 
    \centering
    \begin{center}
    \includegraphics[width=0.46\textwidth]{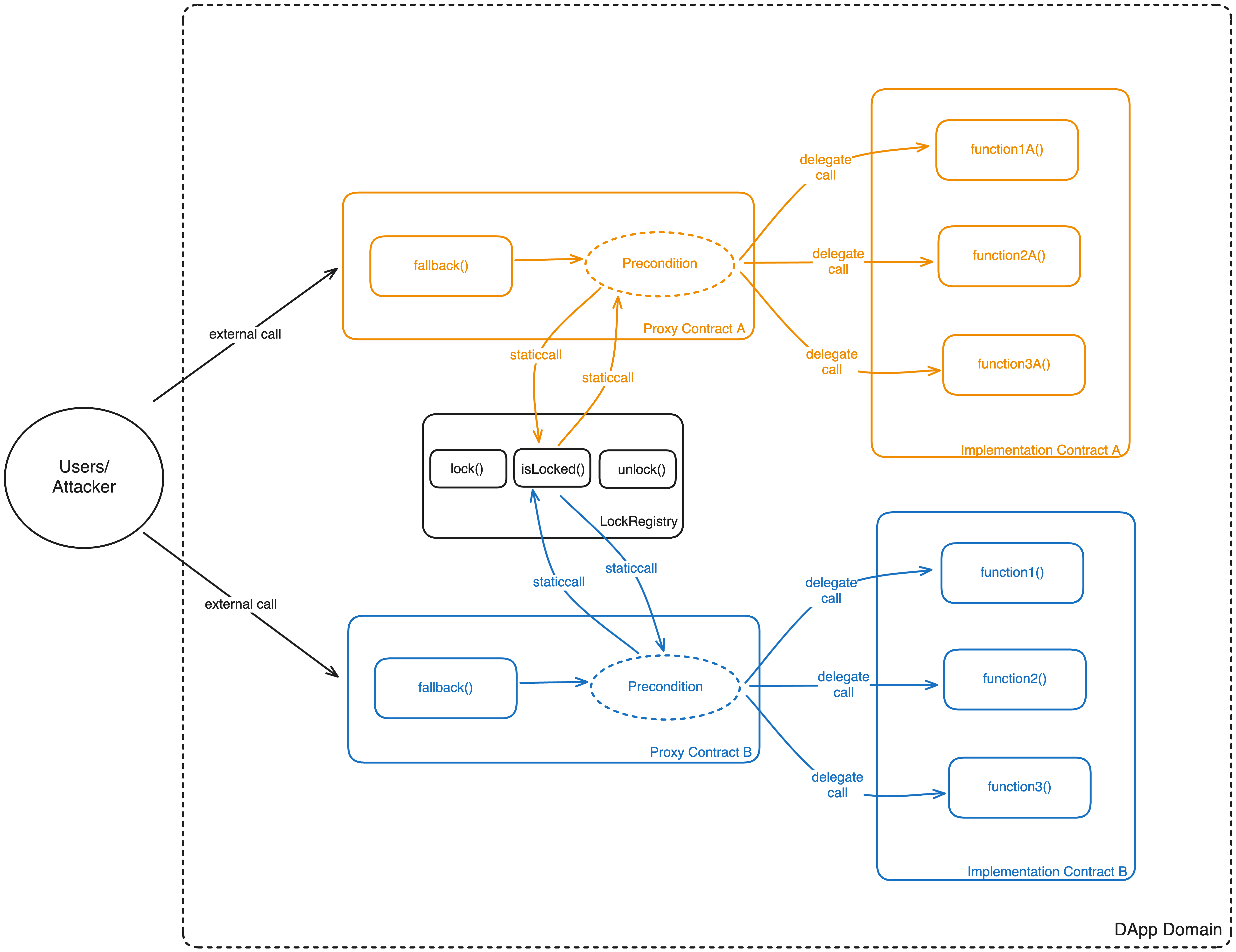}
    \end{center}
    \caption{Architecture Diagram For Sentinel HighSec Mode}
\end{figure}
As demonstrated in Algorithm \ref{alg:high_security_guard}, the pre-execution validation process begins when the \_highSecurityGuard() function retrieves the DOMAIN\_ID\_SLOT, a bytes32 identifier unique to a particular logical domain or contract set, along with the LockRegistry contract address from the proxy's dedicated storage. Sentinel then performs an external call to query the global lock status for the specified domain from the LockRegistry. If the isLocked() function returns true, indicating an active lock, the transaction immediately reverts. When the domain is confirmed as unlocked, the lock activation phase proceeds with an external call to establish the lock for the specified domainId and records msg.sender, which corresponds to the SentinelProxy address, as the designated locker. Also, the STATIC\_CALL\_ALLOWED\_SLOT within the LockRegistry is simultaneously set to false to prevent ROR attacks during the active high-security lock period. This new state of LockRegistry becomes the context for all the SentinelProxy instances and other compatible contracts within the same execution domain. This is followed by the execution of the implementation logic in the proxy's context and return of the delegatecall. The post-execution cleanup phase is similar to optimized mode, which initiates upon successful return of the delegatecall from the implementation contract, resetting the locks. This mode effectively prevents sophisticated reentrancy attacks involving multiple contracts. The global nature of the lock, managed by the LockRegistry, provides coordinated defence against advanced reentrancy attacks that can bypass single-contract mutex implementations due to scope restrictions.

\section{Evaluations}
\begin{table*}[h]
\centering
\caption{Comparison of Security Coverage of Different Security Model Classified According to Reentrancy Types}
\label{etab}
\resizebox{\textwidth}{!}{%
\begin{tabular}{l|c|c|c|c}
\toprule
 & Single Function & Cross Function & Cross Contract & Total \\
\midrule
Sentinel & \progressbar{38}{38} & \progressbar{20}{20} & \progressbar{12}{12} & \progressbar{70}{70} \\[0.3cm] 
OZ ReentrancyGuard & \progressbar{38}{38} & \progressbar{13}{20} & \progressbar{0}{12} & \progressbar{51}{70} \\[0.3cm]
LiqGuard & \progressbar{38}{38} & \progressbar{7}{20} & \progressbar{0}{12} & \progressbar{45}{70} \\[0.3cm]
\bottomrule
\end{tabular}
}
\end{table*}
In this section, we first present the dataset used in the evaluation and introduce our experimental setup of Sentinel. Then we show the evaluation results of Sentinel. And we chose OpenZeppelin's ReentrancyGuard \cite{21} and approach described in \cite{24} (subsequently referred to as LiqGuard) for comparisons. 
\subsection{Dataset}
For the evaluation of Sentinel, our dataset comprises 70 vulnerable smart contracts. The focal objective of the formation of this dataset is to ensure both real-world relevance and broad coverage of reentrancy attack patterns. The dataset is curated by combining the vulnerable contracts from \cite{14}, known vulnerable real-world contracts through public blockchain explorers \cite{31} and remaining contracts were synthetically generated using the SolidiFI injection tool \cite{30}. The dataset is categorized into three primary reentrancy types: Single-Function Reentrancy, Cross-Function Reentrancy, and Cross-Contract Reentrancy. And both Cross-Function Reentrancy and Cross-Contract Reentrancy subsets include instances of Read-Only Reentrancy.
\subsection{Testing Environment}
All smart contracts, including the SentinelProxy, LockRegistry, and the vulnerable implementation contracts from our dataset, were developed and initially tested using Foundry \cite{32}. I was chosen for its native Solidity testing capabilities, built-in gas reporting, and EVM tracing features, which allowed for precise analysis of execution flow and state changes during attack simulations. Local testing was performed on a local Anvil instance (Foundry's local EVM development blockchain), ensuring a consistent, isolated, and deterministic environment without the variability of public testnets. To further validate the practical applicability and observe real-world gas costs, a subset of critical test cases was also deployed and executed on the Sepolia public testnet. 
\subsection{Security Coverage Analysis}
As a core part of our evaluation, we analyzed the security coverage of these three reentrancy mitigation models. The results, summarized in Table \ref{etab}, provide a comparative assessment of their ability to mitigate different categories of reentrancy attacks across our dataset of 70 vulnerable contracts.

The evaluation demonstrates that Sentinel, our proposed proxy-based solution, achieved a perfect mitigation score across all tested reentrancy attack vectors. It successfully safeguarded all 38 Single-Function Reentrancy vulnerabilities, all 20 Cross-Function Reentrancy vulnerabilities, and all 12 Cross-Contract reentrancy vulnerabilities (and the Read-Only reentrancy instances in Cross-Contract and Cross-Function subsets), culminating in a 100\% overall security coverage. This comprehensive performance validates Sentinel's architectural design, which integrates a dual-mode guard and an external LockRegistry (as detailed in Section \ref{psol}), and satisfies the requirement we proposed earlier for any reentrancy guard design to provide complete protection against both intra-contract and complex inter-contract reentrancy scenarios.

In stark contrast, OpenZeppelin's ReentrancyGuard model exhibited significant limitations, particularly as the attack complexity escalated. While it successfully mitigated all 38 single-function reentrancy vulnerabilities, its effectiveness diminished for cross-function reentrancy attacks, mitigating only 13 out of 20 cases. This partial success in cross-function reentrancy scenarios can be attributed to its design as a function-level modifier and a global shared counter design. This makes the security guarantees provided by the ReentrancyGuard susceptible to the underlying logic of the contract and function-level implementations. In our experiments, we observed that the Reentrancy guard performs rather poorly in multi-step operations because of the lock design, which makes all protected functions mutually exclusive. Furthermore, OpenZeppelin's ReentrancyGuard provided no protection against Cross-Contract reentrancy vulnerabilities, mitigating 0 out of 12 cases. This outcome is inherent to its design as a localized, intra-contract guard that fundamentally lacks a mechanism to coordinate locks across different contract addresses or within a broader execution domain.

LiqGuard's performance followed a similar, albeit more pronounced, pattern of diminishing security coverage with increasing attack complexity. Both OpenZeppelin's ReentrancyGuard and LiqGuard have a function-level protection design, where the LiqGuard lock mechanism depends on the fact that the perceived balance and actual balance of the contract will remain the same. It successfully mitigated all 38 single-function reentrancy attacks, demonstrating basic reentrancy protection. However, its coverage for cross-function reentrancy vulnerabilities was notably lower than OpenZeppelin's, at only 7 out of 20 cases. This reduced efficacy in cross-function scenarios, along with its complete failure against CCr attacks (0 out of 12 cases), is attributed to the unsafe calculation of the perceived balance of the contract (which is essential for the reentrancy check) and limited single contract scope. Similar to OpenZeppelin's ReentrancyGuard, LiqGuard also doesn't provide any security guarantees against ROR instances. Our analysis indicates that LiqGuard's mechanism likely suffers from an inability to correctly track or reset its internal lock state across certain nested or indirect call patterns involving two cooperating attackers.
Overall, the results align with the requirements we inferred earlier for the reentrancy mitigation tools to provide a comprehensive security coverage against all reentrancy attack patterns. Sentinel’s proxy-based model provides enhanced implementation-agnostic protection against the full spectrum of reentrancy attack vectors.

\subsection{Gas Consumption}
The gas cost associated with smart contract operations is a critical performance metric, directly impacting user experience and economic viability. Our evaluation quantifies the gas overhead introduced by the SentinelProxy in its different operational modes. The overhead, which remains consistent regardless of the complexity of the underlying implementation contract due to our proxy-based design, stems from specific, predictable EVM operations. For staticcall, the proposed model introduces an overhead of approximately 6400 to 7500 gas, thereby ensuring minimal gas loss for read-only operations. In the optimized mode, the reentrancy lock introduces a primary overhead of approximately 25,000 gas. The High-Security Mode, which involves an external CALL to the ILockRegistry, incurs a slightly higher overhead of 53,000 gas. Considering that an average DeFi (decentralized finance) transaction typically consumes between 200,000 and 500,000 gas, the overhead of our Optimized Mode represents a negligible 5\% to 12.5\% increase, while the High-Security Mode's overhead is a modest 10\% to 25\%. In our experiments, we also observed that the gas overhead for OpenZeppelin's ReentrancyGuard module and LiqGuard primarily depends on the implementation and execution path of the contract, typically ranging from 1,500 to 18,000 gas (only over single contract implementations). This highlights that our proxy, while introducing a predictable overhead, delivers significantly enhanced security coverage, including critical cross-contract and ROR protection, for a competitive gas cost compared to existing, more limited solutions.

\subsection{Analysis of Detection Tools}
While the current paper primarily focuses on providing a reentrancy prevention solution, the majority of the existing efforts are concentrated on identifying the potential vulnerability upfront in the development phase (or pre-deployment phase). However, a direct, like-for-like comparison of Sentinel (a runtime prevention mechanism) with these detection frameworks is not viable because of the fundamental disparity in their methodologies and objectives. While the detection framework aims to identify vulnerabilities, our solution aims to eliminate the attack vector at the execution layer. And the critical limitation of detection-only approaches is that they leave the protection cycle incomplete. While they can pinpoint potential weaknesses, they do not provide a direct method to resolve these issues, nor do they effectively identify novel or evolving attack patterns in complex contract systems.
To highlight this crucial gap, we subjected our dataset of 70 vulnerable contracts to analysis by commonly used reentrancy detection tools. We utilized Mythril \cite{25}, Oyente \cite{26}, and Slither \cite{18}. The results, consistent with findings from other studies \cite{27, 28}, yielded varying detection (true positive) rates by Slither, Mythril, and Oyenete of 71.42\%, 64.28\%, and 57.14\%, respectively. These results underscore a crucial fact: even leading detection frameworks are not entirely foolproof. They can suffer from false negatives (failing to detect actual vulnerabilities), especially when faced with complex or novel reentrancy patterns.

\section{Related Work}
\subsection{Reentrancy Prevention Frameworks}
One of the most common practices to prevent the potential re-entrancy attack is for the developers to use the Checks-Effects-Interactions (CEI) Pattern \cite{33} during the development phase of the target smart contract. This methodology enforces a strict ordering of operations to mitigate reentrancy vulnerabilities. This particular approach is very effective in terms of execution cost due to its simpler approach of undermining the attack vector. However, the limitation of the approach lies in the security coverage, as it assumes that updating limited state variables is sufficient, but complex systems have interdependent states. The CEI pattern provides a necessary but not sufficient condition for reentrancy safety in smart contracts. It guarantees local function safety but does not address system-wide consistency in complex multi-contract architectures. Similar to CEI, Dong et. al. propose a condition-oriented solution that prevents the adversary from performing a reentrancy attack under the same transaction in \cite{34}. When an attacker reenters the transfer function in the smart contract, the reentrant attack is prevented by controlling the state variable. Another innovative approach was proposed in \cite{24}, which calculates the difference between the contract balance and the total balance of all participants in a smart contract before and after any operation in a transaction that changes its state. ReentrancyGuard\cite{21} by Open Zeppelin includes a global function that is shared between all the functions and guards the target contract from the reentrancy attack through the locking approach. Still, this approach has its limitations which were evident from the hacks on the smart contract on which the solution has already been implemented, including but not limited to Fei-Rari (\$80 Million) \cite{35}, Cream (\$18.8 Million) \cite{36}, Orion (\$3 Million) \cite{37}, and BurgerSwap (\$7.2 Million) \cite{38}.

\subsection{Reentrancy Detection Frameworks}
Although detection approaches differ fundamentally from our prevention-oriented methodology, these frameworks are relevant to our work as they establish the theoretical foundation for understanding reentrancy attack vectors and vulnerability characteristics. Methods like Mythril \cite{25}, Oyente \cite{26}, and Manticore \cite{39} provide analysis of vulnerabilities associated with the target smart contract based on a symbolic code execution approach. While these symbolic execution tools are powerful for analyzing smart contracts, they struggle with larger, more complex programs due to the combinatorial explosion of execution paths that need to be explored. Similar to this, Maian \cite{13} also uses symbolic analysis, but its goal is to identify a series of calls that create traces that lead to vulnerabilities. This modification expands the scope of detectable vulnerabilities but still fails to address the fundamental issue of path accessibility problems inherent to symbolic execution models. Slither \cite{18} leverages static analysis to detect reentrancy vulnerabilities. It translates the smart contract into intermediate representation (SlitherIR) and performs analysis on it. But this methodology restricts the scope of analysis, reducing the overall efficiency of this method. In \cite{40}, Liu et al. propose ReGuard, leveraging fuzz testing to detect reentrancy vulnerabilities. It first translates smart contracts to C++ and then performs fuzz testing on smart contracts by iteratively generating random but diverse transactions. Based on the runtime traces, ReGuard further dynamically identifies reentrancy vulnerabilities. Li et al. \cite{27} proposed another fuzzing framework to detect reentrancy vulnerabilities. It preprocesses the source code to limit the problem space while not compromising on the overall scope and associated information.
\section{Conclusion}
Reentrancy attacks remain a persistent and costly vulnerability for decentralized applications. As we've shown, traditional reentrancy guards and detection-focused tools are insufficient, often failing to address the full spectrum of reentrancy vulnerabilities, especially in complex, cross-contract scenarios. To overcome these limitations, this paper introduced Sentinel, a novel proxy-based reentrancy guard. We formalized reentrancy attacks with a precise mathematical model, which provided the foundation for our system's design. Our approach decouples security from the core business logic, integrating a dual-mode guard directly into the proxy layer. This architecture ensures comprehensive protection against a wide array of reentrancy attacks—including sophisticated cross-contract and read-only reentrancy, while also being upgrade-proof. Our evaluation on a dataset of 70 vulnerable smart contracts confirmed Sentinel's effectiveness. The results demonstrate a 40\% higher success rate in preventing attacks compared to existing solutions. Sentinel offers a definitive shift in smart contract security, moving beyond pattern-based detection to a robust, deployable, and provably more effective prevention mechanism. This work establishes a new standard for reentrancy mitigation and provides a strong foundation for future research into type-agnostic security for DApps.

\bibliographystyle{ieeetr}  
\bibliography{references} 
\end{document}